\documentclass[epj]{webofc}
\usepackage[utf8]{inputenc}
\usepackage[varg]{txfonts}   
\usepackage{booktabs}
\usepackage{xcolor}
\definecolor{darkred}{rgb}{0.4,0.0,0.0}
\definecolor{darkgreen}{rgb}{0.0,0.4,0.0}
\definecolor{darkblue}{rgb}{0.0,0.0,0.4}
\usepackage[bookmarks,linktocpage,colorlinks,
    linkcolor = darkred,
    urlcolor  = darkblue,
    citecolor = darkgreen]{hyperref}
\usepackage{subfigure}
\wocname{EPJ Web of Conferences}
\woctitle{Lattice2017}
\newcommand{\In}{\mathrm{in}}
\newcommand{\Out}{\mathrm{out}}
\newcommand{\FV}{\mathrm{FV}}
\newcommand{\alf}{\alpha_{\mathrm{EM}}}

\begin{document}
%
\selectlanguage{english}
\title{%
Including electromagnetism in $K\to\pi\pi$ decay calculations
}
\author{%
\firstname{Norman} \lastname{Christ}\inst{1}\fnsep
\thanks{Speaker, \email{nhc@phys.columbia.edu}, supported in part  by US DOE grant \#DE-SC0011941.} \and
\firstname{Xu} \lastname{Feng}\inst{2,3,4}
}
\institute{%
Department of Physics, Columbia University, New York, NY 10027, USA
\and
School of Physics and State Key Laboratory of Nuclear Physics and
Technology, Peking University, Beijing 100871, China
\and
Collaborative Innovation Center of Quantum Matter, Beijing 100871, China
\and
Center for High Energy Physics, Peking University, Beijing 100871, China
}
\abstract{%
Because of the small size of the ratio $A_2/A_0$ of the $I=2$ to $I=0$ $K\to\pi\pi$ decay amplitudes (the $\Delta I=1/2$ rule) the effects of electromagnetism on $A_2$ may be a factor of 20 larger than given by a naive $O(\alf)$ estimate.  Thus, if future calculations of $A_2$ and epsilon$^\prime$/epsilon are to achieve 10\% accuracy, these effects need to be included.  
Here we present the first steps toward including electromagnetism in a calculation of the standard model $K\to\pi\pi$ decay amplitudes using lattice QCD.}
\maketitle
\section{Introduction}\label{intro}

The current theoretical result for the standard model prediction of the Re($\varepsilon'/\varepsilon$)~\cite{Bai:2015nea}, the measure of direct CP violation in $K\to\pi\pi$ decay, carries a combined statistical and systematic error which is 40\% of the measured value for this quantity.  As reported in Ref.~\cite{Kelly:345}, the RBC and UKQCD Collaborations are working to substantially reduce this error, possibly to the level of 20\%.  Below this level of precision, the effects of electromagnetism on Re($\varepsilon'/\varepsilon$) may become important.  For most processes, electromagnetic corrections are on the order of $\alf \sim 1\%$.  However, the Re($\varepsilon'/\varepsilon$) involves with equal weight the amplitudes $A_0$ and $A_2$ for the decay of a $K$ meson into two pions in the isospin zero and isospin 2 states, respectively.  Since $A_2$ is suppressed relative to $A_0$ by a factor of 22 (the $\Delta I=1/2$ rule), the electromagnetic modifications to $A_0$ can in principle induce corrections to $A_2$ which are 22 times larger than this usual 1\% scale.  Thus, a first-principles, lattice QCD calculation of the two-pion decay of the kaon which includes electromagnetism (and the mass difference between the up and down quark) will soon become important.  Such effects have been extensively studied using chiral perturbation theory~\cite{Cirigliano:1999ie, Cirigliano:1999hj, Cirigliano:2000zw, Wolfe:2000rf, Cirigliano:2003gt} and it would be valuable to verify and extend this work using the methods of lattice QCD.

There are a number of important challenges that need to be addressed in a calculation of these electromagnetic and $m_u-m_d$ effects using lattice methods:
\begin{itemize}
\item The calculation of $K\to\pi\pi$ depends heavily on the finite volume method of Lellouch and L\"uscher~\cite{Lellouch:2000pv} which relies on the exponentially finite-range interactions of QCD.   Adding electromagnetism (E\&M) introduces long-range interactions, inconsistent with the Lellouch-L\"uscher strategy.  This problem is dramatically represented by the fact that the scattering phase shifts which play a central role in the finite-volume treatment of Lellouch and L\"uscher are not even defined when E\&M effects are included, with the long-distance wave functions acquiring phases which grow logarithmically with distance, $\sim\eta\ln(kr)$ where $k$ is the center of mass (CoM) momentum of the scattering particles, $r$ the CoM separation of the outgoing particles and $\eta = Me^2/(4\pi k)$ with $\pm e$ their charge and $M$ their mass.
\item The usual treatment of $K\to\pi\pi$ decay relies on isospin symmetry to distinguish two independent $\pi\pi$ final states, one with $I=0$ and the other with $I=2$.  Electromagnetism and $m_u-m_d$ break isospin symmetry, allowing the $\pi\pi$ states with $I=0$ and $I=2$ to mix.  As a result the final state scattering that is part of the $K\to\pi\pi$ decay becomes a coupled, two-channel problem, requiring a more complex treatment of both the finite- and infinite-volume decay process.
\item A process such as $K\to\pi\pi$ decay which involves the acceleration of charge will contain well-known infrared singularities~\cite{Bloch:1937, Lee:1964is} which are removed by a careful treatment of the possible, near-degenerate final states which include the intended $\pi\pi$ state as well as states with one more more emitted photons in addition to the two pions.  While the effects of such soft radiation can be computed using standard methods for the case of the infinite-volume decay, possible photon emission in a finite-volume lattice calculation may introduce serious complications making the already challenging two-channel problem described above into a problem with three or more channels, including possible three-particle channels.
\end{itemize}

A possible strategy to deal with this series of challenges is to adopt a gauge for the E\&M field that will allow us to separate these issues.  In particular, we propose to work in Coulomb or radiation gauge in which the E\&M vector potential $\vec A$ is required to be transverse; $\vec \nabla \vec A=0$.  Since in a lattice calculation we will naturally work in a finite volume in which the kaon is at rest, no added difficulties are introduced by the choice of such a  Lorentz non-covariant gauge.  In the familiar continuum, Minkowski-space theory, the Lagrangian for the electromagnetic field in Coulomb gauge can be written
\begin{equation}
L_{\mathrm{EM}} = \frac{1}{2} \int d^3 r \left\{ \left(\partial_t \vec A(r)\right)^2 - \left(\vec \nabla \times \vec A(r)\right)^2 + \vec j(r)\cdot\vec A(r) \right\} -\frac{1}{2} \int d^3 r d^3 r^\prime \rho(r)\frac{1}{4\pi|\vec r - \vec r\,^\prime|}\rho(r^\prime),
\label{eq:coulomb}
\end{equation}
a standard, textbook result~\cite{Srednicki:2007qs} for the quantum treatment of the electromagnetic field.  Here $\vec j(r)$ and $\rho(r)$ are the current and charge density operators for the quarks to which the E\&M field couples.  Since we plan to compute the E\&M corrections to $K\to\pi\pi$ decay to first order in $\alf$, the two interaction terms on the right-hand side of Eq.~\eqref{eq:coulomb} can be treated independently and the results simply added together in the end.  This will allow us to consider separately the Coulomb interaction with its long-range distortion of the two-particle scattering problem and the interaction $\vec j\cdot\vec A$ of the transverse photons which requires the Bloch-Nordsiek treatment~\cite{Bloch:1937}.  In this study we will focus on the  corrections arising from the second, Coulomb interaction term and the quark mass difference $m_u-m_d$.

\section{Direct CP violation in $K\to\pi\pi$ decay  without isospin symmetry}

Before discussing possible strategies for computing these effects using lattice QCD, we should examine the effect of electromagnetic interactions on the relation between the parameter $\varepsilon^\prime$ describing direct CP violation in $K\to\pi\pi$ decay and the matrix elements which produce the decay.  The usual definition of $\varepsilon$ and $\varepsilon^\prime$ is general and retains its validity after E\&M effects are included:
\begin{equation}
\eta_{+-} = \frac{^\Out\langle\pi^+\pi^-|H_W|K_L\rangle}{^\Out\langle\pi^+\pi^-|H_W|K_S\rangle} \equiv \varepsilon+\varepsilon^\prime
\quad
\eta_{00} = \frac{^\Out\langle\pi^0\pi^0|H_W|K_L\rangle}{^\Out\langle\pi^0\pi^0|H_W|K_S\rangle} \equiv \varepsilon-2\varepsilon^\prime,
\label{eq:eta-def}
\end{equation}
where the superscript ``out'' indicates a scattering state whose outgoing part contains the particles indicated with a phase chosen to be the same as in the non-interacting case.

Just as in the case of isospin conservation, we can use CPT symmetry to express $\varepsilon^\prime$ in terms of the $\pi\pi$ scattering phase shifts and two decay amplitudes $A_s^\gamma$, $s=0,2$ that will be real if the effective weak operator $H_W$ is CP conserving.  Here $A_s^\gamma$ is a decay matrix element which now includes the effects of E\&M and $m_u \ne m_d$:
\begin{equation}
^\Out\langle(\pi\pi)_s^\gamma|H_W|K^0\rangle = e^{i\delta_s^\gamma}A^\gamma_s, \quad s=0,2.
\label{eq:decay-EM}
\end{equation}
In order to define the quantities which appear in Eq.~\eqref{eq:decay-EM}, we begin with the charged, in and out scattering states $|(\pi\pi)_c\rangle^{\In/\Out}$ with $c= +-$ and $00$, and define the $2\times 2$ unitary scattering matrix $S$:
\begin{equation}
^\Out\langle(\pi\pi)_{c^\prime}(E')|(\pi\pi)_c(E)\rangle^\In \equiv S_{c^\prime c}\,\delta(E'-E), 
\end{equation}
where the two-pion states have angular momentum zero and are normalized to a delta function in energy.  Next we introduce the matrix $\Omega$ which diagonalizes $S$:
\begin{equation}
\left(\Omega^\dagger S \Omega\right)_{s^\prime s} = \delta_ {s^\prime s}\, e^{2i\delta_s^\gamma},
\label{eq:delta-def}
\end{equation}
a relation which also defines the $\pi\pi$ scattering phase shifts $\delta_s^\gamma$ which appear in Eq.~\eqref{eq:decay-EM}.  

The states $|(\pi\pi)_s^\gamma\rangle^\Out$ in Eq.~\eqref{eq:decay-EM} are the two combinations of $|(\pi\pi)_{+-}\rangle^{\Out}$ and $|(\pi\pi)_{00}\rangle^{\Out}$ which diagonalize $S$:
\begin{equation}
|(\pi\pi)_s^\gamma \rangle^{\Out/\In}
            = \sum_{c=+-,00} \Omega_{c,s} |(\pi\pi)_c\rangle^{\Out/\In}.
\end{equation}

Here $|(\pi\pi)_s^\gamma\rangle^\Out$ are now eigenstates of the $2\times 2$, $\pi\pi$ scattering matrix which includes the effects of isospin breaking.  While these states are no longer eigenstates of isospin, we expect them to be close to isospin eigenstates and therefore label each of them with the isospin, $I=0$ and 2, of the corresponding, nearby isospin eigenstate.  (To avoid ambiguity, we add the superscript $\gamma$ when identifying a quantity which includes the effects of E\&M and $m_u \ne m_d$ with a symbol conventionally used for the isospin conserving case.)  The time reversal symmetry of QCD allows us to choose the matrix $\Omega$  to be real:
\begin{equation}
\Omega = \left(\begin{array}{cc} \cos{\theta^\gamma} & \sin{\theta^\gamma} \\
                                                    -\sin{\theta^\gamma} & \cos{\theta^\gamma}
                                      \end{array} \right)
              = \left(\begin{array}{cc} \sqrt{2/3} & \sqrt{1/3} \\ -\sqrt{1/3} & \sqrt{2/3}
                                      \end{array} \right) + (\mbox{isospin breaking}),
\label{eq:theta-def}
\end{equation}
where the matrix on the right-hand side is constructed from the usual Clebsch-Gordan coefficients relating the isospin eigenstates to those with particles carrying definite charges.
The matrix $\Omega$ can be used to express the observed, charged states that appear in Eq.~\eqref{eq:eta-def} in terms of the amplitudes $A_s^\gamma$ and phases $\delta_s^\gamma$:
\begin{equation}
^\Out\langle(\pi\pi)^\gamma_c|H_W|K^0\rangle 
                          = \sum_{s=0,2}\Omega_{cs} e^{i\delta^\gamma_s}A^\gamma_s.
\label{eq:charged_amplitude}
\end{equation}

Finally we can generalize the conventional expression for $\varepsilon^\prime$ to this isospin broken case if Eq.~\eqref{eq:charged_amplitude} is substituted into Eq.~\eqref{eq:eta-def} and the result expanded to first order in $A_2^\gamma/A_0^\gamma$:
\begin{equation}
\varepsilon^\prime = \frac{1}{3}(\eta_{+-}-\eta_{00})
= \frac{\sin 2\theta}{\sin 2\theta^\gamma}\,\frac{i e^{i(\delta_2^\gamma-\delta_0^\gamma)}}{\sqrt{2}}
            \frac{\mathrm{Re}A_2^\gamma}{\mathrm{Re}A_0^\gamma}
  \left(\frac{\mathrm{Im}A_2^\gamma}{\mathrm{Re}A_2^\gamma}
            - \frac{\mathrm{Im}A_0^\gamma}{\mathrm{Re}A_0^\gamma}\right)
\label{eq:epsilon-prime-EM}
\end{equation}
where $\theta =35.26^\circ$, the value of $\theta^\gamma$ when the effects of E\&M and $m_u\ne m_d$ are removed.

\section{Choosing the finite volume formulation}

Next we must adopt a finite volume formulation for the $K\to\pi\pi$ decay which can be used in a lattice QCD calculation and related to the infinite volume amplitudes of interest. Of particular interest here is the treatment of the long-range Coulomb potential.  An approach that is frequently adopted, referred to as QED$_L$ replaces the function $1/r$ by a periodic function obtained by writing $1/r$ as a Fourier series and discarding the Fourier mode with $\vec k = (0,0,0)$.  While such an approach may be convenient for a calculation in which an extrapolation to infinite volume, $L\to\infty$ is to be taken, it is not appropriate for the present calculation where we must exploit finite volume quantization to create a physical final state.  In our case terms of order $1/L$ play an important role in the calculation and cannot be neglected.  Most of the unwanted effects fall exponentially with the volume and the few unphysical effects which fall as a power of the box size $L$ are explicitly calculated and removed~\cite{Lellouch:2000pv}.
Such a removal of finite volume effects would be difficulty for a  QED$_L$ approach in which the potential itself has large $1/L$ distortions even for $r$ close to zero.

We propose to adopt a different, truncated finite-volume Coulomb potential in which the potential is unchanged for $r \le R_T$ and set to zero for $r>R_T$:
\begin{equation}
V_T(r) = \left\{\begin{array}{cl} \frac{e^2}{4\pi r} & r \le R_T \\ 0 & r > R_T \end{array}\right. .
\end{equation}
Here the truncation radius $R_T$ should be chosen to lie outside of the region in which the two charged pions interact but smaller than the $L/2$.  Thus, we propose to compute $\pi\pi$ scattering and the $K\to\pi\pi$ decay using this truncated Coulomb potential.  

As discussed below, the lattice calculation is performed in finite volume and generalizations~\cite{He:2005ey, Hansen:2012tf} of the results L\"uscher~\cite{Luscher:1990ux} and Lellouch and L\"uscher~\cite{Lellouch:2000pv} are used to determine the infinite volume scattering and decay amplitudes.  The truncated Coulomb potential is ideally suited to these methods which assume that all interactions have a finite range.  However, the infinite-volume results obtained in this way will correspond to the truncated potential $V_T(r)$ not the physical $1/r$ potential.   Since $V_T(r)$ differs from the physical potential only when $r$ is sufficiently large that the particles being studied are no longer affected by the strong interactions, we expect that the corrections needed to relate the truncated potential results to those from the physical Coulomb potential can be computed in perturbation theory.  

For the case of non-relativistic potential scattering this is straight-forward to demonstrate.  Since the truncated and Coulomb problems agree for $r<R_T$, the wave functions in that region will agree and the correct Coulomb wave function can be determined by extending the $r \le R_T$ results to $r > R_T$ by matching the $V_T(r)$ wave function to a solution to the Coulomb scattering problem at $r=R_T$.  For the case of the relativistic field-theory problem the calculation of these needed connections has not yet been worked out.

This relation between the truncated and physical Coulomb potential results for $s$-wave scattering and decay is complicated by the infinite range of the Coulomb potential.   This problem is conventionally avoided by using a screened Coulomb potential $e^{-r/R_S}/r$ where $R_S$ is the screening radius which can be chosen to be much larger than the hadronic scale and represents the behavior of the material in which the decay or scattering experiment is being performed.   For $\pi\pi$ scattering, the $s$-wave phase shift is not well-defined and if a partial wave description of $\pi\pi$ scattering is desired the effects of screening need to be explicitly included.  For the problem of determining $\varepsilon^\prime$ this problem is absent because the amplitude $\eta_{+-}$ is defined as the ratio of two amplitudes for decay into the $\pi^+\pi^-$ final state.  The effects of the screening mechanism cancel between the numerator and denominator and our proposed calculation with the truncated Coulomb potential should also require no correction.

\section{Lattice QCD calculation}

\begin{figure}
\centering
\includegraphics[width=8cm,clip]{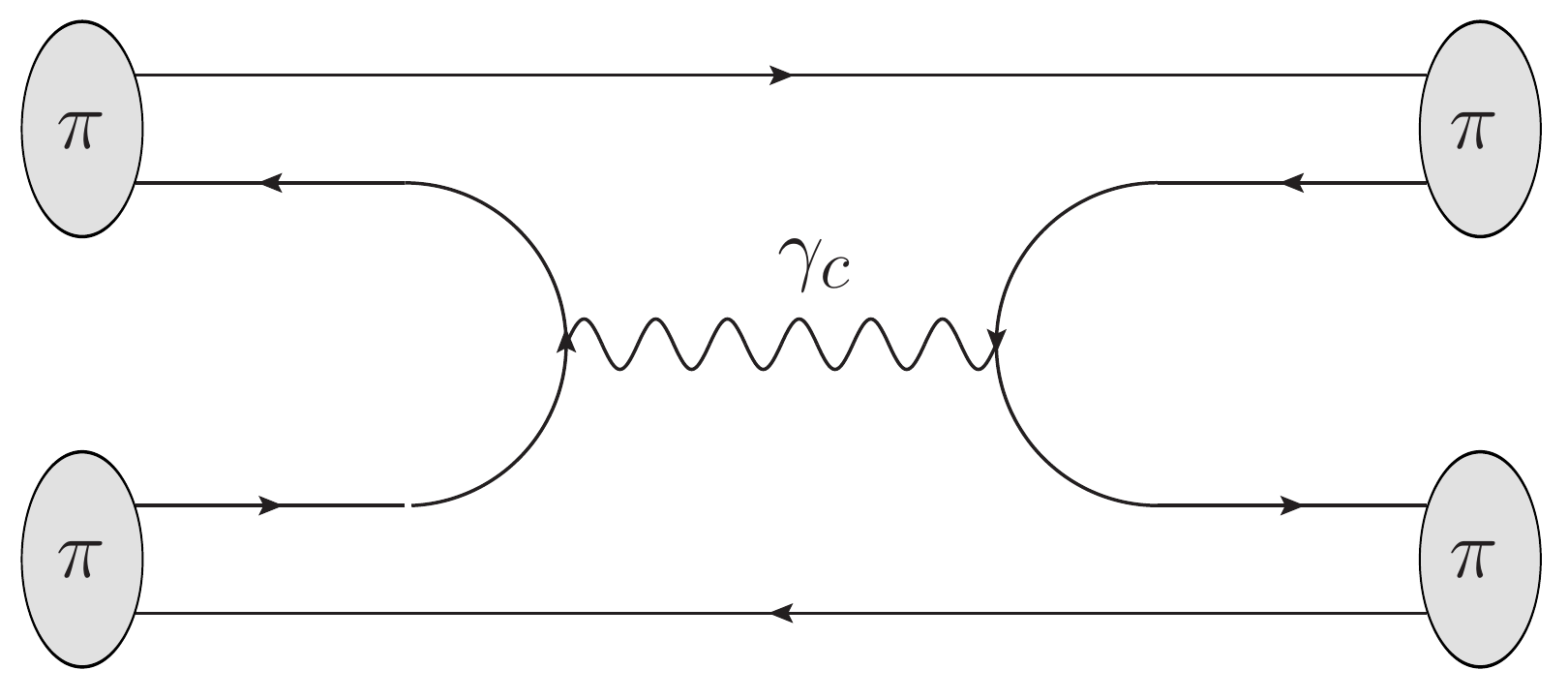}
\caption{Example of a $\pi\pi$ scattering diagram in which the Coulomb interaction appearing in Eq.~\protect{\eqref{eq:coulomb}} enters in first order.}
\label{fig:decay-graph}
\end{figure}

We next briefly discuss the lattice QCD calculation that would support the approach outlined above.  At least in principle, it should be straight-forward to add the instantaneous Coulomb interaction given in Eq.~\eqref{eq:coulomb} to the complete set of isospin conserving contractions that were evaluated to obtain the result presented in Ref.~\cite{Bai:2015nea}.   An example diagram is shown in Figure~\ref{fig:decay-graph}.  Thus, we propose to explicitly expand in $\alf$ and in the isospin breaking, light-quark mass term, evaluating all possible single insertions of the Coulomb potential and this mass term into the contractions that contribute to $\pi\pi$ scattering and $K\to\pi\pi$ decay.  

The results of such a calculation of $\pi\pi$ scattering can be described by examining the set of $\pi\pi$-$\pi\pi$ correlators obtained from the two-pion interpolating operators $(\pi\pi)_I(t)$ constructed to have a definite isospin I and acting at the time $t$.  The form of the result can be obtained by inserting the set of low-lying, finite-volume energy eigenstates that will appear in such an isospin violating calculation and expanding that result in powers of $\alf$ and $m_u-m_d$:
\begin{eqnarray}
\bigl\langle 0\bigl|(\pi\pi)_I(t) (\pi\pi)_{I'}(0)\bigr|0\bigr\rangle
&=& \sum_{s=0,2}\bigl\langle 0\bigl|(\pi\pi)_I(t)\bigr|(\pi\pi)_s^{\gamma,\FV}\bigr\rangle
\,e^{-E_s^{\gamma,\FV} t}\,
             \bigl\langle (\pi\pi)_s^{\gamma,\FV}\bigl|(\pi\pi)_{I'}(0)\bigr|0\bigr\rangle \\
&=& \left(\begin{array}{cc}
e^{-E_0^{\FV,0}t}\Bigl(1+2N_{00}^{(1)}+E_0^{\FV,1} t\Bigr) & e^{-E_0^{\FV,0}t}N_{20}^{(1)}+e^{-E_2^{\FV,0}t}N_{02}^{(1)} \\
e^{-E_0^{\FV,0}t}N_{20}^{(1)}+e^{-E_2^{\FV,0}t}N_{02}^{(1)} & e^{-E_2^{\FV,0}t}\Bigl(1+ 2N_{22}^{(1)}+E_2^{\FV,1} t\Bigr) 
\end{array}\right)_{I,I'}. 
\label{eq:Oalpha-corr}
\end{eqnarray}
Here $E_I^{\FV,0}$, $I=0$, 2 are the finite volume energies of the $I=0$, and 2 states to zeroth order in isospin violating effects while the $E_I^{\FV,1}$ are the first-order shifts in those energies that result from the addition of E\&M and $m_u \ne m_d$ effects: $E^{\gamma,\FV}_I = E^{\FV,0}_I + E_I^{\FV,1} + \mathcal{O}(\alf^2)$.  The $2\times 2$ matrix $N^{(1)}$ describes the overlap between the $\pi\pi$ interpolating operators $(\pi\pi)_I(t)$ and the finite-volume, energy eigenstates $\bigr|(\pi\pi)_s^{\gamma,\FV}\bigr\rangle$ determined to first order in $\alf$ and $m_u-m_d$.  We have normalized the $\pi\pi$ interpolating operators $(\pi\pi)_I(t)$ so that 
\begin{equation}
\bigr\langle 0\bigl|\pi\pi_I(0)\bigr|(\pi\pi)_s^{\gamma,\FV}\bigr\rangle = \delta_{Is}+N_{Is}^{(1)}
\label{eq:norm}
\end{equation}
and assumed the conventional charge-conjugation symmetry for those operators.

Previous calculations in the absence of isospin breaking can be used to impose the normalization condition in Eq.~\eqref{eq:norm} and determine the energies $E_I^{\FV,0}$.  By comparing lattice results with the expression given in Eq.~\eqref{eq:Oalpha-corr} the first-order energy shifts $E_I^{\FV,1}$ and the finite-volume, first-order mixing coefficients $N_{II'}^{(1)}$ can be determined.  These mixing coefficients can then be used to construct interpolating operators $(\pi\pi)^\gamma_s$ which will create the finite-volume energy eigenstates, accurate to first order in $\alf$ and $m_u-m_d$:
\begin{equation}
(\pi\pi)^\gamma_s(t) =   (\pi\pi)_s(t) - \sum_I N_{s,I}^{(1)}\, (\pi\pi)_I(t). 
\end{equation}
These operators can be used to obtain the finite-volume, stationary-state version of the matrix elements that appear in Eq.~\eqref{eq:decay-EM} from the two three-point functions:
\begin{equation}
\bigl\langle (\pi\pi)^\gamma_s(t_{\pi\pi})\, H_W(t_{\mathrm{op}})\, K(t_K)\bigr\rangle \quad s=0,2\; ,
\end{equation}
where these matrix elements are to be evaluated to first order in $\alf$ and $m_u-m_d$.

The discussion in this section is intended to make the proposed lattice calculation more concrete and to make explicit the information that can be obtained.  Specifically we will be able to determine the two finite-volume energy eigenvalues $E_s^{\gamma,\FV}$ and the corresponding energy eigenstates $|(\pi\pi)_s^{\gamma,\FV}\rangle$ for $s=0$ and 2.  At a minimum we will need to perform calculations with two different boundary conditions: one chosen to give $E_0^{\FV,0} = M_K$ and a second designed so the $E_2^{\FV,0} = M_K$.  If these calculation are performed as was done in the isospin conserving case, we would use anti-periodic boundary for the $d$ quark in two directions to achieve $E_2^{\FV,0} = M_K$~\cite{Blum:2011ng} and G-parity boundary conditions in all three direction for the $E_0^{\FV,0}=M_K$ case~\cite{Bai:2015nea}, using a $32^3\times 64$ ensemble with $1/a=1.38$ GeV.

\section{Finite volume corrections}

We will now discuss the method to be used to relate the finite-volume energies and matrix elements that can be determined in a lattice QCD calculation to the infinite-volume phase shifts $\delta_s^\gamma$ and matrix elements needed in Eq.~\eqref{eq:epsilon-prime-EM} to calculate $\varepsilon^\prime$ in the standard model accurate to first order in $\alf$ and $m_u-m_d$.  This general two-channel decay problem has been analyzed in detail by Hansen and Sharpe~\cite{Hansen:2012tf} and we will rely on their results in this discussion.  (The first analysis of two-channel finite-volume energy quantization was carried out in Ref.~\cite{He:2005ey} but we focus on the presentation of Hansen and Sharpe because they also discuss the finite volume calculation of two-channel particle decay.)

In summarizing their results, we will refer to the example system that they present where the two channels are $\pi\pi$ and $KK$ states in an $s$-wave with isospin 0.  Since the pion and kaon masses are unequal, the $\pi\pi$ and $KK$ channels are easily distinguished and a single finite-volume quantization condition relates the energy of a finite-volume energy eigenstate to the two infinite-volume scattering phase shifts and a single rotation angle relating the eigenstates of the scattering matrix to the more physically accessible  $\pi\pi$ and $KK$ states.  In our case these would correspond to our two phase shifts $\delta_0^\gamma$ and $\delta_2^\gamma$ and the rotation angle $\theta^\gamma$ of Eqs.~\eqref{eq:delta-def} and \eqref{eq:theta-def}.  For our case we write this quantization condition as 
\begin{equation}
\Phi_\alpha\bigl(E,\delta^\gamma_0(E),\delta^\gamma_2(E), \theta^\gamma(E) \bigr) = 0.
\end{equation}
Here the parameter $\alpha$ identifies the volume and boundary conditions which determine the finite volume problem being studied.  This condition suggests that we must perform three calculations with three volumes and boundary conditions $\alpha_i$, $i=1$, 2, 3 adjusted so that for each case $E=M_K$.  The three independent conditions $\Phi_{\alpha_i}\bigl(E,\delta^\gamma_0(E),\delta^\gamma_2(E), \theta^\gamma(E) \bigr) = 0$ can then be solved to determine the three parameters describing $\pi\pi$ scattering at the kaon mass: $\delta_0^\gamma$, $\delta_2^\gamma$ and the rotation angle $\theta^\gamma$.

While this approach is required for the $\pi\pi$-$KK$ problem studied by Hansen and Sharpe, our case is different in two respects.  First, we are expanding to first order in $\alf$ and $m_u-m_d$ and at zeroth order two independent channels can be easily identified by their isospin.  Second, the two channels that are to be distinguished when considering the actual scattering or decay problems are the states $(\pi\pi)_c$ for $c=+-$ and $00$, {\it i.e.} the $\pi^+\pi^-$ and $\pi^0\pi^0$ states.  From the point of view of scattering the zeroth-order $I=0$ and $I=2$ states are ``non-degenerate'', with scattering phase shifts which are quite different.  Thus, when isospin breaking is introduced the new, mixed states will be related to the $\pi^+\pi^-$ and $\pi^0\pi^0$ states by an angle $\theta^\gamma$ which differs from the zeroth-order angle $\theta=35.26^\circ$ given by the usual Clebsch-Gordan coefficient by a small amount of order $\alf$ or $m_u-m_d$.  This implies that to first order in our expansion parameters $\alf$ and $m_u-m_d$ the function $\Phi_\alpha\bigl(E,\delta^\gamma_0(E),\delta^\gamma_2(E), \theta^\gamma(E) \bigr)$ will not depend on $\theta^\gamma$ since in the finite volume calculation $\theta^\gamma$ is meaningful only when the $\pi^+\pi^-$ and $\pi^0\pi^0$ states can be distinguished kinematically by having distinct masses.  Thus, $\theta^\gamma$ must always appear in the product $\theta^\gamma (m_{\pi^+} - m_{\pi^0})$.  The term of interest, $(\theta^\gamma - \theta)(m_{\pi^+} - m_{\pi^0})$ is second order and will not enter our first-order lattice calculation.  This absence of $\theta^\gamma-\theta$ from the quantization condition simplifies the proposed lattice QCD calculation since now only two sets of volumes and boundary conditions are needed to determine the two scattering phase shifts $\delta_0^\gamma$ and $\delta_2^\gamma$.  

The discussion above explains how a pair of finite volume calculations can be used to determine the two phase shifts $\delta_s^\gamma(M_K)$, $s=0$, 2 that are needed to evaluate Eq.~\eqref{eq:epsilon-prime-EM}.  As shown in Ref.~\cite{Hansen:2012tf} the infinite volume decay amplitudes $A_s^\gamma$ (the quantities $v_1$ and $v_2$ in that paper) can be determined by computing the decay matrix elements using the finite-volume scattering eigenstates $|(\pi\pi)_s^{\gamma,\FV}\rangle$ and applying Eqs.~(86)-(92) of that paper. If these seven equations are expanded to first order in $\alf$ and $m_u-m_d$, all the needed quantities can be obtained from the two finite-volume $\pi\pi$ scattering calculations described above except for derivatives of the phase shifts with respect to energy which could be determined from two additional $\pi\pi$ scattering calculations at energies close to $M_K$.  

However, the relation between the infinite volume $\pi\pi$ scattering eigenstates and the physical charged or neutral infinite volume states, $|(\pi^+\pi^-)^\gamma\rangle^\Out$ and $|(\pi^0\pi^0)^\gamma\rangle^\Out$ to first order in $\alf$ and $m_u-m_d$ does require knowledge of the first-order difference $\theta^\gamma - \theta$ which has not been determined.  This enters Eq.~\eqref{eq:epsilon-prime-EM} for $\epsilon^\prime$ as an uncertainty in the angle $\theta^\gamma$.  While our lack of knowledge of $\theta^\gamma$ will lead to an uncertainty in $\epsilon^\prime$ of order $\alf$ and $m_u-m_d$ this may be viewed as a minor difficulty affecting the result for $\epsilon^\prime$ on the order of 1\%.  Recall that our target is the much larger effect of isospin breaking that may affect the cancellation of the terms in the large curved brackets in Eq.~\eqref{eq:epsilon-prime-EM}, possibly at the 10\% level.

\section{Conclusion}

The proposal presented here represents the first of a number of steps that will be required to carry out a first-principles calculation of the important isospin breaking contribution to the standard model prediction for $\varepsilon'$.  The two evident further barriers to such a calculation are the complexity of the calculation proposed and the remaining component of the Coulomb-gauge E\&M interaction given in Eq.~\eqref{eq:coulomb}, which has not been discussed here.   

In concept it appears straight-forward to add the instantaneous Coulomb interaction appearing in Eq.~\eqref{eq:coulomb} into the 350 separate contractions which contribute to the $K\to\pi\pi$ calculation with G-parity boundary conditions presented in Ref.~\cite{Bai:2015nea}.  However, in practice automated techniques may be needed to handle the several thousand contractions that must be evaluated.  These techniques may be similar to those that will be required for other frontier lattice QCD calculations in particle and nuclear physics.  The difficulties implied by the second component of the Coulomb-gauge interactions, the explicit coupling to massless, transverse photons are also serious.  However, these effects need only be evaluated in lowest order perturbation and the most difficult, long-range parts are known from classical physics so it may be possible to find a computational scheme that avoids at least some of these difficulties for those parts of the problem which require lattice QCD.

\end{document}